\documentclass[sigconf]{acmart}

\AtBeginDocument{%
  }

\setcopyright{acmlicensed}
\copyrightyear{2025}
\acmYear{2025}
\setcopyright{cc}
\setcctype{by}
\acmConference[CIKM '25]{Proceedings of the 34th ACM International Conference on Information and Knowledge Management}{November 10--14, 2025}{Seoul, Republic of Korea}
\acmBooktitle{Proceedings of the 34th ACM International Conference on Information and Knowledge Management (CIKM '25), November 10--14, 2025, Seoul, Republic of Korea}\acmDOI{10.1145/3746252.3761390}
\acmISBN{979-8-4007-2040-6/2025/11}

\usepackage{graphicx}
\usepackage{multirow}
\usepackage{makecell}
\usepackage{stfloats}
\usepackage{makecell}
\usepackage{algorithm, algpseudocode}
\usepackage{amsmath}
\usepackage{array}
\usepackage{subcaption}
\usepackage{multirow}
\usepackage{graphicx}
\usepackage[normalem]{ulem}
\useunder{\uline}{\ul}{}
\begin{document}
\nocite{*}

\title{Efficient Multimodal Streaming Recommendation via Expandable Side Mixture-of-Experts}

\author{Yunke Qu}
\affiliation{%
  \institution{The University of Queensland}
  \city{Brisbane}
  \state{}
  \country{Australia}
}
\email{yunke.qu@uq.net.au}

\author{Liang Qu}
\affiliation{
    \institution{Edith Cowan University}
    \city{Perth}
    \state{}
    \country{Australia}
}
\email{l.qu@ecu.edu.au}

\author{Tong Chen}
\affiliation{%
  \institution{The University of Queensland}
  \city{Brisbane}
  \state{}
  \country{Australia}
}
\email{tong.chen@uq.edu.au}

\author{Quoc Viet Hung Nguyen}
\affiliation{
  \institution{Griffith University}
  \city{Gold Coast}
  \state{}
  \country{Australia}
}
\email{henry.nguyen@griffith.edu.au}

\author{Hongzhi Yin}
\authornote{Corresponding author}
\affiliation{%
  \institution{The University of Queensland}
  \city{Brisbane}
  \state{}
  \country{Australia}
}
\email{h.yin1@uq.edu.au}

\renewcommand{\shortauthors}{Yunke Qu, Liang Qu, Tong Chen, Quoc Viet Hung Nguyen, \& Hongzhi Yin}

\begin{abstract}
Streaming recommender systems (SRSs) are widely deployed in real-world applications, where user interests shift and new items arrive over time. As a result, effectively capturing users' latest preferences is challenging, as interactions reflecting recent interests are limited and new items often lack sufficient feedback. A common solution is to enrich item representations using multimodal encoders (e.g., BERT or ViT) to extract visual and textual features. However, these encoders are pretrained on general-purpose tasks: they are not tailored to user preference modeling, and they overlook the fact that user tastes toward modality-specific features such as visual styles and textual tones can also drift over time. This presents two key challenges in streaming scenarios: the high cost of fine-tuning large multimodal encoders, and the risk of forgetting long-term user preferences due to continuous model updates.

To tackle these challenges, we propose Expandable Side Mixture-of-Experts (XSMoE), a memory-efficient framework for multimodal streaming recommendation. XSMoE attaches lightweight side-tuning modules consisting of expandable expert networks to frozen pretrained encoders and incrementally expands them in response to evolving user feedback. A gating router dynamically combines expert and backbone outputs, while a utilization-based pruning strategy maintains model compactness. By learning new patterns through expandable experts without overwriting previously acquired knowledge, XSMoE effectively captures both cold start and shifting preferences in multimodal features. Experiments on three real-world datasets demonstrate that XSMoE outperforms state-of-the-art baselines in both recommendation quality and computational efficiency. Code is available at https://github.com/qykcq/Efficient-Multimodal-Streaming-Recommendation-via-Expandable-Side-Mixture-of-Experts.

\end{abstract}



\maketitle

\section{Introduction}

Streaming recommender systems (SRSs) represent a common setting in real-world applications such as e-commerce platforms and social media, where user preferences shift and new items arrive over time. To keep up with these changes, SRSs continuously update in response to user interactions, enabling real-time personalization without the delay and computational overhead of full model retraining \cite{zhang2023incremental, schnabel2022situating}. While capable of timely updates, most existing SRSs still rely solely on ID-based user-item interactions \cite{guo2019streaming, qiu2020gag, mi2020ader, wang2018streaming, latifi2022streaming, li2020time, peng2021learning, yin2023multi}. These methods often struggle with data sparsity and limited generalization, as interactions reflecting users’ latest preferences are limited and newly introduced items frequently lack sufficient feedback.

A common solution to address the above limitations is to incorporate multimodal side information, such as item images and textual descriptions, which enrich item representations by incorporating semantic features extracted from associated visual and textual modalities. This motivates the use of multimodal recommenders \cite{liang2023mmmlp, guo2024lgmrec}, which typically extract features using pretrained encoders such as ViT \cite{dosovitskiy2021image} for visual content and BERT \cite{devlin2019bert} for text. While these models offer strong semantic representations, they are pretrained on general-purpose tasks and are not optimized to capture user-specific preferences in recommendation scenarios. In practice, users often exhibit fine-grained and evolving tastes toward modality-specific attributes—for example, shifting from intricate visual designs to minimalist aesthetics, or from factual descriptions to emotionally expressive languages \cite{yu2021visually, orabyetal2015thats}. We refer to this as modality-level preference drift. Addressing it in streaming environments requires updating multimodal encoders over time. This presents two key challenges in streaming scenarios: the high cost of fine-tuning large multimodal encoders, and the risk of forgetting long-term user preferences due to continuous model updates.

The first challenge, efficient adaptation, is particularly difficult in streaming environments characterized by large data volume and velocity. Some prior works attempt to address this by reducing training data via reservoir sampling \cite{guo2019streaming} or hyper-networks that blend in model weights trained on past and present data \cite{zhang2020retrain}. While effective in ID-based settings, these methods are ill-suited for multimodal SRSs, where the training time and memory complexity are dominated by updating pretrained encoders with large parameter sizes.

The second challenge, retention of past knowledge, is critical in SRSs where both short-term and long-term user preferences must be captured \cite{xu2020graphsail, wang2018neural}. In streaming settings, models are trained sequentially on new data. Directly fine-tuning often leads to the forgetting of long-term user preferences \cite{zhang2024influential}. Existing SRSs methods can be broadly categorized into two groups: (1) rehearsal-based methods \cite{guo2019streaming, diaz2012real, chen2013terec, mi2020memory, qiu2020gag, wang2018streaming, mi2020ader} that store and replay old data and (2) knowledge-transfer methods \cite{wang2020practical, mi2020ader, zhang2020retrain, mi2020ader} that transfer old knowledge to the current model \cite{zhang2023incremental}. Rehearsal-based methods require continuous updates to the replay buffer and sampling based on a carefully designed heuristic. Storing and replaying high-dimensional multimodal content is computationally intensive. While current knowledge-transfer approaches partially reduce the training cost by avoiding repeated training on old data via knowledge distillation \cite{mi2020ader, xu2020graphsail} or a hyper-model \cite{zhang2020retrain}, they still need to update every parameter, incurring high computational cost in multimodal SRSs with parameter-heavy multimodal encoders. 

To address the first challenge, efficient adaptation, we propose to leverage side-tuning to achieve both training time and memory efficiency when updating the model with new knowledge. Previous studies \cite{fu2024iisan, sung2020lst, zhang2020side, xu2023san, fu2024dtl, xin2024parameter} have demonstrated that side-tuning achieves both training time efficiency and memory savings compared to full fine-tuning and other Parameter-Efficient Finetuning (PEFT) methods such as Adapter \cite{houlsby2019parameter} and LoRa \cite{hu2022lora}, making it well-suited for efficient adaptation in multimodal SRSs. Specifically, the side-tuning network fine-tunes one or more Transformer blocks with one Feedforward Network (FFN). Unlike traditional PEFT methods that modify the backbone and require storing intermediate activations, side-tuning operates independently of the backbone, eliminating the need for backpropagation through the main network and significantly reducing GPU memory usage and training time \cite{fu2024iisan, sung2020lst, xin2024parameter}. 

Recently, model-editing methods \cite{yu2024boosting, wang2024self, lee2020neural, serra2018overcoming} that integrate additional trainable model components while leaving old parameters untouched have gained popularity in addressing catastrophic forgetting. The Mixture-of-Experts (MoE) architecture is commonly adopted by these methods, where expert specializes in a subset of tasks or data distributions, allowing the model to adapt to new information without overwriting previously acquired knowledge \cite{mu2025moesurvey}. Therefore, to address the second challenge, knowledge retention, we draw inspiration from these methods and generalize each layer in the current side-tuning network, which can be seen as a single-expert MoE, to a multi-expert architecture. In our design, each expert is trained on data from a specific time period. Starting from a single expert, we dynamically expand the MoE at each layer of the side-tuning network to accommodate new data upon the arrival of a new time window. Upon expansion, all existing experts are frozen and a new trainable FFN is added to each layer of the side-tuning network. At each layer, a router dynamically integrates the output of each expert and the frozen backbone encoders using weighted sum, ensuring that each expert contributes proportionally to their relevance to incoming data. We also expand the router by adding a new column when expansion occurs. Compared to using a single trainable expert throughout training, this expandable architecture allows new information to be encoded in newly added experts rather than overwriting old ones, thereby mitigating forgetting.

To further control model growth, we introduce a utilization-based pruning mechanism. In each training epoch, we monitor the contribution of each expert to the final output using a simple norm-based heuristic. At the end of a time window, experts whose relative contribution falls below a threshold are deemed underutilized are pruned along with their corresponding router weights. This ensures that the model remains both scalable and efficient over time.

We present E\textbf{x}pandable \textbf{S}ide \textbf{M}ixture-\textbf{o}f-\textbf{E}xperts (XSMoE) for multimodal SRSs, a memory-efficient framework for multimodal streaming recommendation that combines the strengths of side-tuning for efficient training, expandable MoE for retaining long-term patterns, and expert pruning for bounded model complexity. Our key contributions are as follows:
\begin{enumerate}
\item We propose the first multimodal streaming recommendation framework that integrates visual and textual features via side-tuning.
\item We introduce a dynamically expandable MoE architecture that enables efficient adaptation to new data while preserving past knowledge.
\item We design a utilization-based pruning strategy to maintain memory and computational efficiency by removing underutilized experts.
\item We conduct comprehensive experiments on three real-world datasets, demonstrating that XSMoE's superior performance and efficiency compared to state-of-the-art SRS baselines.
\end{enumerate}

\section{Related Works}
\subsection{Continual Learning}
Continual learning (CL) enables models to learn new information sequentially while retaining previously acquired knowledge, making it essential for adapting to evolving data distributions. Continual learning methods aimed at mitigating catastrophic forgetting are typically categorized into three groups: rehearsal-based approaches \cite{vitter1985random, riemer2018learning, chaudhry2019tiny}, regularization-based strategies \cite{kirk2017overcoming, ahn2019uncertainty, nguyen2018variational}, and model-editing techniques \cite{wang2024self, li2024atlas, lee2020neural, serra2018overcoming, zhang2020side}. Rehearsal-based approaches require continual maintenance of a replay buffer and a carefully designed sampling strategy, which can become computationally prohibitive in multimodal settings due to the cost of storing and replaying high-dimensional content. Regularization-based approaches attempt to preserve previously learned knowledge by constraining parameter updates. However, in SRSs, where the priority is to optimize for future user behavior, merely restricting network changes does not ensure that past knowledge is effectively transferred to address the present task. \cite{zhang2023incremental}. Model-editing techniques either adapt a small subset of parameters \cite{fang2025alphaedi}, introducing nontrivial overhead in identifying them, or dynamically expand the model by freezing existing components and appending new modules \cite{yu2024boosting, wang2024self, lee2020neural}. Our proposed method, XSMoE, belongs to the second kind of model-editing methods. 

\subsection{Streaming Recommendation}
SRSs process continuous streams of user interactions, along with their associated item features, and update the recommender model incrementally to provide personalized recommendations \cite{zhang2023incremental, schnabel2022situating}. Zhang et al. \cite{zhang2023incremental} categorized existing SRSs into two main strategies based on the CL techniques they adopt: rehearsal-based approaches \cite{guo2019streaming, diaz2012real, chen2013terec, mi2020memory, mi2020ader, qiu2020gag, wang2018streaming, mi2020ader} that store and replay old data, and knowledge-transfer approaches \cite{wang2020practical, mi2020ader, xu2020graphsail, zhang2020retrain} that aim to transfer old knowledge to the current model \cite{zhang2023incremental}. Kim et al. \cite{kim2022meta} also proposed to use adaptive learning rate specific to user-item interaction parameter pairs. Our proposed method follows the second strategy.

Despite advancements in streaming recommendation, most existing SRSs remain ID-based, relying solely on user-item interaction histories. ID-based approaches are liable to data sparsity issues and poor generalization. Although SML \cite{zhang2020retrain} is model-agonistic and can be used on multimodal recommenders, it uses a CNN to blend in model weights trained on previous and current data. This practice may incur excessive time and memory complexity in multimodal SRSs due to the multimodal encoders with a massive amount of parameters. Multimodal content integration via efficient updating of the multimodal encoders remains an underexplored direction for enhancing recommendation quality in streaming settings.

\subsection{Multimodal Recommendation}
A multimodal recommender system (MRS) processes multiple types of data sources such as texts \cite{fu2024iisan, liang2023mmmlp, guo2024lgmrec, li2023text, xiong2021cikm, wu2019reviews}, images \cite{fu2024iisan, liang2023mmmlp, guo2024lgmrec, qiu2021casalrec, gupta2020image, packer2018visually}, and locations \cite{yin2015joint, yin2016joint} to generate personalized recommendations by integrating information from different modalities. The pipeline consists of three key stages: raw feature representation, feature interaction, and recommendation \cite{liu2024multimodal}. In the raw feature representation stage, tabular metadata and multimodal inputs (such as images and text) are transformed into numerical representations through modality-specific encoders, such as ViT \cite{dosovitskiy2021image} for visual data and BERT \cite{devlin2019bert} for textual data. In the feature interaction stage, representations from different modalities are fused into a unified embedding space, often using attention mechanisms or graph-based models to capture complex cross-modal relationships. Finally, in the recommendation stage, the fused multimodal representations are used to model user preferences and predict relevant items, typically through ranking-based learning or neural recommendation models \cite{liu2024multimodal}. Despite the advantages of incorporating multimodal content, the computational efficiency of MRSs is significantly constrained by the high cost of running large-scale modality encoders, making efficient adaptation strategies essential for practical deployment. This challenge is especially critical in multimodal SRSs, and our method is designed to address this issue.

\section{Problem Formulation}
Our proposed method operates in a streaming recommendation setting. The task is to predict the next item a user will interact with based on a continuous stream of user-item interaction data. Formally, let $L$ denote the maximal sequence length and \([x_1, x_2, \dots, x_L]\) represent the sequence of items a user interacted with, where \(x_l \in [x_1, x_2, \dots, x_L]\) corresponds to an item interacted with at time step \(l\). In a multimodal setting, each item is associated with an image and textual content (e.g., titles and descriptions), represented as \(v_{x_l}\) and \(t_{x_l}\). Thus, $x_l$ can be denoted as \(x_l = (v_{x_l}, t_{x_l})\), incorporating both visual and textual modalities. We focus on visual and textual content because these modalities are both readily available and widely used in modern recommender systems \cite{zhou2023multimodalsurvey}. 

Given a prefix of item interactions \( [x_1, x_2, \dots, x_L]\) of length $L$ and a historical user-item interaction matrix \(\mathbf{R}\), we generate a ranked list \(y = [y_1, y_2, \dots, y_{|I|}]\), where  \(y_j\)  represents the prediction score for item  \(j\), $\mathcal{I}$ denotes the set of items, and $|\mathcal{I}|$ denotes the number of items. Since recommendation models in a streaming setting must operate  in a time-ordered and continuously evolving manner, the model dynamically updates itself with new data rather than undergoing periodic retraining. As users typically focus on a small set of relevant items , only the  top-k ranked items  from \(y\) are recommended.  

To simulate the streaming setting, we follow the settings of previous SRS \cite{guo2019streaming, qiu2020gag, yin2023multi} works: the dataset \(D\)  is sorted in chronological order and divided into \(T+1\) sequential parts that are mutually exclusive: $D = [D_0, D_1, \dots, D_{T}]$. During the warm-up stage, a newly initialized model is trained on $D_0$. At each time step $s \in [1, T-1]$, the dataset \(D_s\) is further split into a training set  \(D_s^{\text{train}}\) and a  validation set  \(D_s^{\text{val}}\). The model is trained on  \(D_s^{\text{train}}\), validated on  \(D_s^{\text{val}}\), and then evaluated on the next time step’s data \(D_{s+1}\) to assess its ability to generalize to future interactions. 

\section{Methodology}

\subsection{Model Overview}
In a nutshell, XSMoE attaches a lightweight side-tuning network to each frozen pretrained backbone multimodal encoder to enable efficient adaptation to new streaming data. During the warm-up stage, each side-tuning network only has a single FFN expert per layer, as in \cite{fu2024iisan}. As data from a new time window arrives, XSMoE appends a new FFN per layer to incorporate fresh knowledge while previously trained experts remain frozen to preserve past knowledge. Following the MoE architecture, a router integrates the outputs of the experts and the backbone encoder using weighted sum, ensuring smooth adaptation without catastrophic forgetting. To prevent unbounded model growth, we monitor the utilization of each expert using a norm-based heuristic, maintaining both computational efficiency and scalability.  

\subsection{Expandable Side Mixture-of-Experts}
\label{sec:xsmoe}

\begin{figure}
    \centering
    \includegraphics[width=0.85\linewidth]{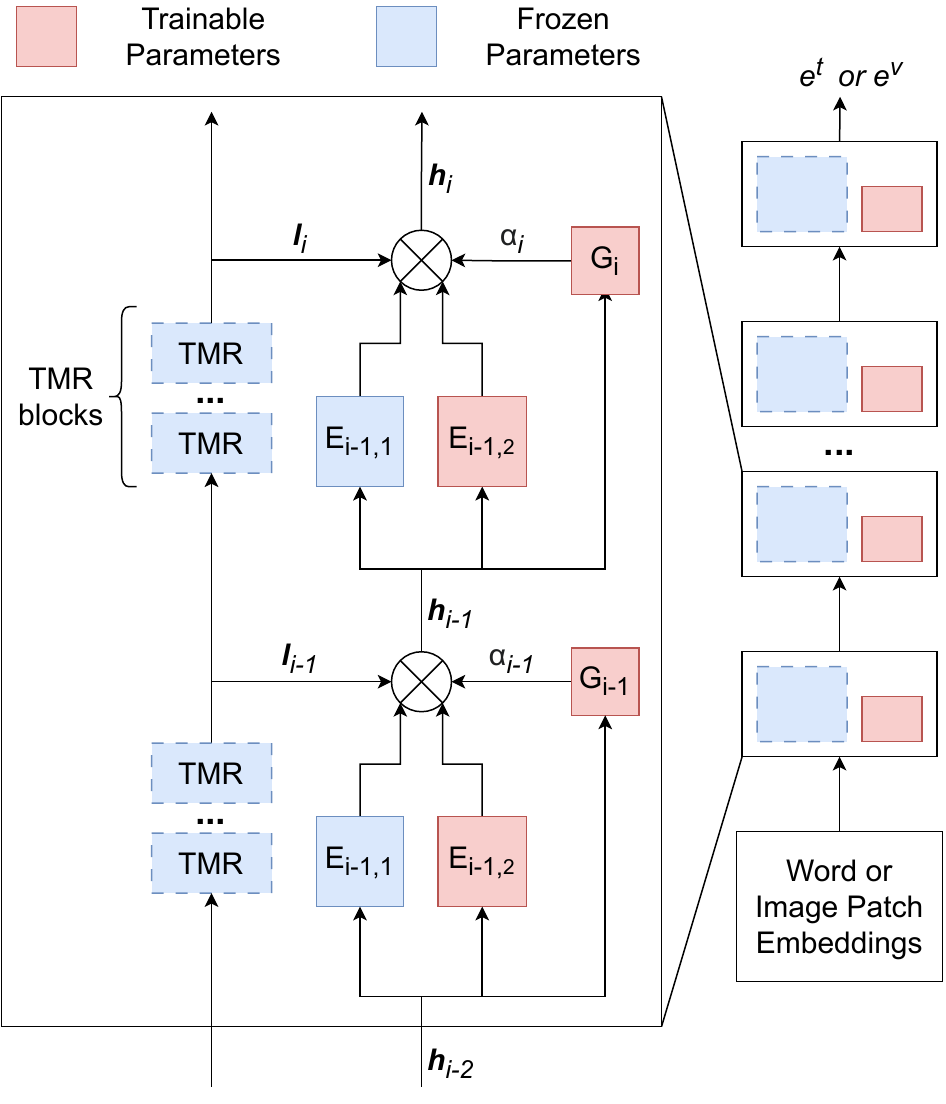}
    \caption{The model structure after one expansion. The visual encoder ViT and the textual encoder BERT are dashed because we use their precomputed outputs. The actual models are never loaded into memory afterwards. TMR denotes a Transformer layer. G denotes the router. In this example, every two Transformer layers are grouped together so the number of layers in the side-tuning network is halved.}
    \label{fig:expansion}
\end{figure}

XSMoE consists of two frozen backbone encoders that extract visual and textual features, along with two modality-specific side-tuning networks that enable efficient adaptation in a streaming environment. The backbones include two pretrained encoders: a textual encoder (i.e., BERT \cite{devlin2019bert}) for textual content and a visual encoder (i.e., ViT \cite{dosovitskiy2021image}) for images. Both backbone encoders have one embedding layer, followed by $M$ hidden Transformer-based layers. We denote the output of the initial embedding layer as $\textbf{l}^{t/v}_0$ and denote the outputs of the Transformer-based layers as $[\textbf{l}^{t/v}_1, \textbf{l}^{t/v}_2, \dots, \textbf{l}^{t/v}_M]$. $t$ and $v$ denote the textual modality and the visual modality, respectively. In the remainder of this paper, we omit the superscripts for modality for simplicity where there is no confusion. $\textbf{l}_0$ is the word embeddings for BERT and image patch embeddings for ViT. $\textbf{l}_M$ is the final sentence-level embeddings for BERT and image-level embeddings for ViT. $[\textbf{l}_0, \textbf{l}_1, \dots, \textbf{l}_M]$ are precomputed and cached. Therefore, the backbone encoders do not need to be loaded into memory. Optionally, to further reduce the computational overhead, the Transformer layers can be evenly grouped to reduce the number of layers in each side-tuning network.

We instantiate two modality-specific light-tuning networks, each of which contains $M$ layers. We denote their outputs as $[\textbf{h}_1, \textbf{h}_2, \dots, \textbf{h}_M]$, one for each $\textbf{l}_i \in [\textbf{l}_1, \textbf{l}_2, \dots, \textbf{l}_M]$. During the warm-up stage, each layer of the side-tuning network has one trainable FFN-based expert. We denote the initial expert at the $i$-th layer as $E_{i,1}$. Upon the arrival of a new time window with new data, XSMoE appends a new expert network to each layer. We denote the set of all experts at the $i$-th layer as $\mathcal{E}_i = \{E_{i, 1}, E_{i,2},\dots,E_{i, N_{i}}\}$, where $N_i$ is the number of experts at the $i$-th layer. Each FFN-based expert $E_{i,j}$ contains a down-projection layer, a $\text{GELU}$ activation function, and an up-projection layer with a skip connection. It receives the output from the previous layer $\textbf{h}_{i-1}$:
\begin{equation}
    E_{i, j}(\textbf{h}_{i-1}) = \textbf{W}_{\text{up}}(\text{GELU}(\textbf{W}_{\text{down}}\textbf{h}_{i-1})) + \textbf{h}_{i-1}  
\end{equation}
For the special case of experts at the first layer, they receive the word embeddings or image patch embeddings, namely $\textbf{h}_0=\textbf{l}_0$:
\begin{equation}
    E_{1, j}(\textbf{h}_0) = E_{1, j}(\textbf{l}_0),
\end{equation}
where $E_{1, j}$ denotes any expert network at the first layer. 

Each layer in the side-tuning network is equipped with a router, implemented as a lightweight FFN that outputs $N_i + 1$ scalar weights—one for the frozen backbone encoder and $N_i$ for the experts. We denote the router at the $i$-th layer as $G_i$. At the $i$-th layer, the router takes as input the output from the previous side-tuning layer $\textbf{h}_{i-1}$ then it combines the output from the frozen backbone encoder $\textbf{l}_i$ and the outputs of all experts $[ E_{i,1}(\textbf{h}_{i-1}), \ldots, E_{i,N_i}(\textbf{h}_{i-1}) ]$ at the same layer. The final output of the layer $\textbf{h}_i$ is computed as a weighted sum:
\begin{align}
\alpha_{i} &= \text{Softmax}(\textbf{W}_i \textbf{h}_{i-1}) \in \mathbb{R}^{N_i + 1}, \nonumber \\
\textbf{h}_i &= \alpha_{i,0} \cdot \textbf{l}_i + \sum_{j=1}^{N_i} \alpha_{i,j} \cdot E_{i,j}(\textbf{h}_{i-1}),
\end{align}
where $\textbf{W}_i$ is the learnable router parameters for layer $i$, and $\alpha_{i} = \{ \alpha_{i,0}, \alpha_{i,1}, \ldots, \alpha_{i,N_i} \}$ are the normalized routing weights computed via a Softmax over the router’s output. Here, $\alpha_{i,0}$ is assigned to the frozen backbone output $\textbf{l}_i$, while the rest correspond to the experts. This soft mixture formulation allows the model to dynamically determine the contribution of each expert and the frozen backbone at every layer, enabling fine-grained control over adaptation and past knowledge use in streaming environments.

To integrate new knowledge without forgetting past information, XSMoE periodically expands its architecture by adding a new expert network per layer in the beginning of every time window. All previous experts are frozen to preserve previous knowledge. To speed up training, the weights of the new expert are initialized to be the average of the weights of all previous experts at that layer. This initialization not only accelerates convergence but also helps avoid expert collapse: scenarios where the newly added expert is assigned zero weights. 

Likewise, at each layer, after an expert is added, a new router column is also appended to the router. All the router columns remain active to adaptively modulate the use of each expert network. Figure \ref{fig:expansion} illustrates the architecture after one expansion. 

After obtaining the latent representations $e^v$ for visual content and $e^t$ for textual content, we  then obtain the item embedding by fusing the two modality representations via a fully-connected layer:
\begin{equation}
    \textbf{e} = \text{FC}([\textbf{e}^v; \textbf{e}^t]),
\end{equation}
where $\text{FC}$ denotes the fully-connected layer and $[;]$ denotes concatenation. 

\subsection{Expert Pruning} \label{sec:pruning}
While the expert expansion strategy in XSMoE allows the model to flexibly adapt to evolving data distributions, it may lead to unbounded growth in the number of experts. Such growth increases memory consumption and model complexity, posing scalability challenges in long-term streaming scenarios. To address this, XSMoE introduces a utilization-aware pruning mechanism that selectively removes underutilized experts, thereby maintaining a compact and efficient model.

At each layer of the side-tuning network, we denote the output of the $j$-th expert as $\textbf{z}_j$, and its corresponding router weight as $\alpha_j$. The utilization score $r_j$ for expert $E_j$ is defined as:
\begin{equation} \label{eq:util}
r_j = \frac{\| \alpha_j \textbf{z}_j \|_2}{\sum_{i=1}^{N} \| \alpha_i \textbf{z}_i \|_2},
\end{equation}
where $\| \cdot \|_2$ is the Euclidean norm and $N$ is the number of experts at that layer. Note that the subscript indexing the layers are omitted for simplicity. The numerator measures the magnitude of expert $E_j$'s contribution to the final output, while the denominator aggregates the total contribution from all expert sources, including the frozen backbone and other experts. As a result, $r_j \in [0, 1]$ quantifies the relative influence of expert $E_j$ in producing the layer's output.

An expert is considered underutilized if its utilization score falls below a predefined threshold $\tau$. The underutilized expert, along with its associated router column, is pruned to control the model size. If multiple experts at the same layer are underutilized at the same time, we only prune the one with the smallest utilization score, ensuring only one removal per layer at a time. Intuitively, if the weighted output magnitude $\| \alpha_j \textbf{z}_j \|_2$ is negligible, its contribution to the final output is minimal. In such cases, pruning these experts causes little to no degradation in performance because they do not play a significant role in decision-making. 

\subsection{Sequence Encoding and Loss Function}
Given the prefix of an item sequence $[x_1, x_2, \dots, x_L]$, we extract the latent representations for each item and obtain a sequence of item embeddings $[\textbf{e}_1, \textbf{e}_2, \dots, \textbf{e}_L]$, which is then fed into a sequence encoder to capture temporal dependencies in user-item interactions. The user embedding is given as:
\begin{equation}
    \textbf{e}_u = \text{SeqEncoder}([\textbf{e}_1, \textbf{e}_2, \dots, \textbf{e}_L]),
\end{equation}
where $\text{SeqEncoder}(\cdot)$ is the Transformer-based sequence encoder. Given an input embedding sequence of length $L$, the encoder first adds learned positional embeddings and applies layer normalization and dropout. It then passes the result through stacked Transformer blocks, each comprising a multi‐headed self‐attention layer followed by a position‐wise feed‐forward network and residual connections with layer normalization.

Given a user $u$ and an item $i$, the model predicts a relevance score $y_{ui}$, which is the dot product of the sequential encoder output and the item embedding:
\begin{equation}
    y_{ui} = \textbf{e}_i \cdot \textbf{e}_u
\end{equation}

We optimize XSMoE using the In-Batch Debiased Cross-Entropy Loss as in \cite{fu2024iisan}:
\begin{align} \label{eq:objective}
    D_{ui} &= \exp(y_{ui} - \log(p_i)) + \sum_{\substack{j \in [B] \\ j \notin \mathcal{I}_u}} \exp(y_{uj} - \log(p_j)), \nonumber \\ 
    \mathcal{L}_{\text{CE}} &= - \sum_{u \in \mathcal{U}} \sum_{i\in \mathcal{I}} \log \left( \frac{\exp(y_{ui} - \log(p_i))}{D_{ui}} \right),
\end{align}
where $\mathcal{U}$ denotes the user set, $\mathcal{I}$ denotes the item set,  \(p_i\) is the popularity of item \(i\), \(\mathcal{I}_u\) is the set of items interacted with user \(u\), and \(|B|\) denotes the size of the data batches that are indexed by $j$.

\begin{algorithm}[t]
\caption{Expandable Side MoE for Multimodal SRSs (XSMoE)}
\label{alg:xsmoe}
\begin{algorithmic}[1]
\Require Streaming data $D = [D_0, D_1, D_2, \dots, D_T]$

\State Obtain precomputed results from backbone encoders $[\textbf{l}_0, \textbf{l}_1, \dots, \textbf{l}_M]$
\State Initialize one side-tuning network with $M$ layers per modality
\State Warm up the model on $D_0$
\For{$s = 1, \dots, T$} 
    \State Obtain new dataset $D_s = \{ D_s^{\text{train}}, D_s^{\text{val}} \}$
    \State Update model on $D_s^{\text{train}}$ w.r.t. Equation \eqref{eq:objective}
    \State Test model on next time step $D_{s+1}$ w.r.t. HR@10 and NDCG@10
    \State Compute utilization scores $[r_1, r_2, \dots, r_N]$ w.r.t. Equation \eqref{eq:util} for each expert at each layer 
    \State Freeze all experts at each layer
    \State Initialize a new expert $E_{N+1}$ at each layer 
    \State Add a new router column to the router at each layer 
    \State Prune underutilized experts
\EndFor
\end{algorithmic}
\end{algorithm}

The overall procedure is summarized in Algorithm \ref{alg:xsmoe}.

\subsection{Complexity Analysis}
Hereby we analyze the computational complexity of XSMoE w.r.t. parameter efficiency, training time complexity, and GPU memory efficiency per modality, following the protocols in \cite{fu2024iisan}. We denote the input and output size of each expert as $d$, the number of layers in each side-tuning network as $M$, the down-projection size as $d'$, and the number of experts at each layer as $N$. Typically, we have $N < d' < d$ due to the expert pruning mechanism. For simplicity, we assume each layer contains the same number of experts $N$.

\subsubsection{Parameter Efficiency}
Each layer of XSMoE contains two components: $N$ FFN experts and 1 router. Each expert contains one up-projection layer and one down-projection layer, both of which has $dd'$ parameters. Therefore, the parameter size of an expert is $2dd'$. The input size of the router is $d$ and the output size is $N+1$. Hence, each router has $dN+d$ parameters. Therefore, given $M$ layers, each with $N$ experts and one router, the number of parameters of XSMoE is $M(2Ndd'+Nd+d) \approx \mathcal{O}(MNdd')$. 

However, at any point of the training stage, only one expert, along with the router, remains trainable per layer. Therefore, the number of trainable parameters is $M(2dd'+Nd+d) \approx \mathcal{O}(Mdd')$. 

\subsubsection{Training Time Complexity}
The per-epoch training time complexity is dominated by three parts: (1) forward passes, (2) backward passes, and (3) weight updates \cite{fu2024iisan}. (1) involves all trainable and frozen parameters, and is thus $\mathcal{O}(MNdd')$. (2) and (3) only involve trainable parameters and are thus $\mathcal{O}(Mdd')$. Therefore, the overall per-epoch training time complexity is $\mathcal{O}(MNdd')$. However, per-epoch training time complexity does not translate to overall training time complexity. We will empirically show that overly simplifying a model may slow down convergence and as a result, the model may need more epochs to converge in Section \ref{sec:ablation}.

\subsubsection{GPU Memory Complexity}
The GPU memory usage is dominated by (1) model weights, (2) gradients, (3) optimizer states, and (4) activations \cite{fu2024iisan}. At each layer, (1) is $\mathcal{O}(Ndd')$. (2) only involves trainable parameters and is thus $\mathcal{O}(dd')$. The optimizer states of Adam double the trainable parameter count are thus $\mathcal{O}(2dd')$. The activations are stored after forward passes and are used for backpropagation. In XSMoE, activations are stored for the router and the experts. At each layer, the router that computes Softmax scores $N$ experts and its corresponding backbone encoder. Hence, the activations for the routers are $\mathcal{O}(N+1)$. For $N$ experts, the activations are $\mathcal{O}(Nd)$. In summary, the GPU memory complexity of XSMoE is $\mathcal{O}(Ndd') + \mathcal{O}(dd') + \mathcal{O}(2dd') + \mathcal{O}(N+1) + \mathcal{O}(Nd) \approx \mathcal{O}(Ndd')$ for each layer and $\mathcal{O}(MNdd')$ for $M$ layers. 

\section{Experiments}
In this section, we aim to answer the following research questions:
\begin{itemize}
    \item RQ1: How does XSMoE perform compared to existing ID-based SRSs?
    \item RQ2: Is fine-tuning multimodal encoders necessary in multimodal SRSs?
    \item RQ3: Does the expandable MoE architecture alleviate catastrophic forgetting and improve recommendation performance?
    \item RQ4: Can expert pruning improve training time, parameter, and GPU memory efficiency while preserving recommendation performance?
    \item RQ5: How does each modality contribute to the final recommendation performance?
\end{itemize}

\subsection{Experimental Setup}
\subsubsection{Datasets}
We evaluate XSMoE and the baselines on three real-world datasets: Amazon Home and Kitchen (Amazon Home), Amazon Electronics \cite{ni2019justifying}, and H\&M \footnote{https://www.kaggle.com/competitions/h-and-m-personalized-fashion-recommendations/data}, all of which are commonly used in multimodal recommendation research \cite{fu2024iisan, yuan2023where}. The two Amazon datasets have reviews between May 1996 - Oct 2018. The original H\&M dataset contains 33,265,846 records from 2019 to 2022 but we only select data from 2022. We sort each dataset in chronological order and divide it into 10 parts, each of which contains an equal number of records. For the Amazon datasets, we concatenate the product title and the product description to incorporate more textual information. Then for each part, we apply $k$-core processing, with $k=5$ for the two Amazon datasets and $k=10$ for the H\&M dataset. After processing, the Amazon Home dataset has 2,220,678 interactions between 257,475 users and 126,312 items; the Amazon Electronics dataset has 2,464,044 records between 266,457 users and 121,186 items; the H\&M dataset has 4,152,911 interactions between 170,071 users and 53,310 items. The selected datasets are large enough to evaluate the proposed method within a resource-constrained environment. Within each part, the first 85\% of interactions are used for training, while the remaining 15\% serve as the test set. We deem all the observed interactions as positive and unobserved interactions as negative. By leveraging datasets from different domains and platforms, we can comprehensively evaluate XSMoE across diverse recommendation scenarios.

\subsubsection{Baselines} 
Since there has been no work on multimodal SRS, we select baseline methods from three ID-based SRSs: 
\begin{itemize}
   \item SSRM \cite{guo2019streaming}: a streaming session-based recommendation method that enhances attention-based user intent modeling with matrix factorization and incorporates a reservoir-based memory with active sampling to efficiently handle high-velocity session data.
   \item GAG \cite{qiu2020gag}: a streaming session-based recommendation method that builds global-attributed session graphs to jointly model user and session representations, and uses a Wasserstein reservoir to retain representative historical data for continual adaptation.
   \item GIUA-GNN \cite{yin2023multi}: a streaming session-based recommendation model that integrates global user and item embeddings with local session information using an attentional graph neural network, and enhances feature extraction with a bi-directed attentional graph convolutional network.
\end{itemize}
We also select three multimodal recommender systems originally designed for static, non-streaming scenarios: 

\begin{itemize}
    \item MMMLP \cite{liang2023mmmlp}: an efficient MLP-based sequential recommender that models multi-modal user interaction sequences using layered feature mixing and fusion.
    \item IISAN \cite{fu2024iisan}: IISAN is a decoupled parameter-efficient side-tuning framework for multimodal sequential recommendation that performs both intra- and inter-modal adaptation, significantly improving training speed and GPU memory efficiency while matching the performance of full fine-tuning and other PEFT methods.
    \item LGMRec \cite{guo2024lgmrec}: LGMRec is a multimodal recommender that jointly models local and global user interests by decoupling collaborative and modality-based signals via a local graph module and enriching them with global user/item dependencies through a hypergraph embedding module.
\end{itemize}

To evaluate the performance of existing continual learning (CL) approaches in the streaming setting, we enhance the above static multimodal recommenders with two representative model-agnostic CL techniques: reservoir sampling as in \cite{guo2019streaming, qu2024scall} and Knowledge Distillation (KD) loss. Reservoir sampling is a rehearsal-based method that maintains a fixed-size buffer of historical item sequences. During training, the model is updated using both current and replayed samples from the buffer. KD loss, on the other hand, is a regularization-based approach where the current model is penalized for deviating from the predictions of a reference model trained on earlier data. Specifically, we minimize the Kullback-Leibler divergence between the current model’s output distribution and that of the model from the past time window. We use the subscripts “RH” and “KD” to denote these two enhanced versions of the base models. 

\begin{table*}[t] 
\caption{The recommendation performance w.r.t. HR@10 and NDCG@10 in percentage (\%) across different time windows on the Amazon Home, Amazon Electronics (abbreviated as Amazon Elec.), and H\&M datasets. $[D_2, \dots, D_9]$ denote the test data chunks. For XSMoE, $\tau$ is set to 0.15, 0, and 0 on the Amazon Home, Amazon Electronics, and H\&M datasets, respectively. The best results are highlighted.}
\label{tab:overall}
\setlength{\tabcolsep}{1.5pt}
\resizebox{\textwidth}{!}{%
\begin{tabular}{cccccccccccccccccccc}
\Xhline{2\arrayrulewidth}

\multirow{2}{*}{Datasets} & \multirow{2}{*}{Methods} & \multicolumn{2}{c}{$D_2$} & \multicolumn{2}{c}{$D_3$} & \multicolumn{2}{c}{$D_4$} & \multicolumn{2}{c}{$D_5$} & \multicolumn{2}{c}{$D_6$} & \multicolumn{2}{c}{$D_7$} & \multicolumn{2}{c}{$D_8$} & \multicolumn{2}{c}{$D_9$} & \multicolumn{2}{c}{AVG} \\ \cline{3-20} 

&  & HR & NDCG & HR & NDCG & HR & NDCG & HR & NDCG & HR & NDCG & HR & NDCG & HR & NDCG & HR & NDCG & HR & NDCG \\ \hline

\multirow{10}{*}{\begin{tabular}[c]{@{}c@{}}Amazon\\ Home\end{tabular}} & SSRM & 9.29 & 6.66 & 8.81 & 6.51 & 8.60 & 6.54 & 7.28 & 5.42 & 8.20 & 6.29 & 8.48 & 6.58 & 8.42 & 6.51 & 6.29 & 4.71 & 8.17 & 6.15 \\ 

& GAG & 12.31 & 9.89 & 10.92 & 8.84 & 10.23 & 8.32 & 8.63 & 6.85 & 9.27 & 7.48 & 9.70 & 7.92 & 9.42 & 7.65 & 6.97 & 5.33 & 9.68 & 7.79 \\

& GIUA-GNN & 15.53 & 13.32 & 13.29 & 11.26 &  12.19& 10.46 & 9.81 & 8.25 & 10.79 & 9.21 & 10.94 & 9.27 & 10.76 & 9.14 & 7.54 & 6.22 & 11.36 & 9.64 \\ 

& $\text{MMMLP}_{\text{RH}}$ & 13.56 & 11.26 & 11.57 & 9.52 & 10.80 & 8.84 & 8.78 & 7.35 & 9.40 & 7.82 & 10.09 & 8.31 & 9.76 & 8.23 & 6.72 & 5.65 & 10.09 & 8.37 \\  

& $\text{LGMRec}_{\text{RH}}$ & 12.19 & 9.71 & 10.33 & 8.29 & 9.21 & 7.34 & 7.32 & 5.87 & 7.90 & 6.45 & 8.45 & 6.99 & 8.38 & 7.05 & 5.77 & 4.83 & 8.69 & 7.07 \\ 

& $\text{IISAN}_{\text{RH}}$ & 21.45 & 17.67 & 19.03 & \textbf{15.45} & 16.61 & 13.47 & 13.67 & 10.78 & 14.60 & 11.69 & 14.41 & 11.30 & 13.90 & 10.85 & 10.32 & 7.65 & 15.50 & 12.38 \\

& $\text{MMMLP}_{\text{KD}}$ & 11.20 & 8.80 & 11.33 & 9.10 & 10.19 & 8.22 & 8.58 & 6.96 & 8.84 & 7.24 & 9.58 & 8.04 & 9.33 & 7.94 & 6.62 & 5.58 & 9.45 & 7.73 \\ 

& $\text{LGMRec}_{\text{KD}}$ & 10.84 & 8.33 & 9.49 & 7.42 & 9.04 & 7.12 & 6.80 & 5.31 & 7.20 & 5.75 & 7.52 & 6.03 & 8.23 & 6.81 & 5.37 & 4.36 & 8.06 & 6.39 \\

& $\text{IISAN}_{\text{KD}}$ & 20.65 & 16.92 &  17.55 & 14.30 & 15.49 & 12.66 & 12.61 & 10.22 & 13.42 & 10.90 & 13.34 & 10.73 & 12.63 & 10.16 & 9.36 & 7.21 & 14.38 & 11.64 \\

& XSMoE & \textbf{22.67} & \textbf{18.02} & \textbf{19.43} & 15.37 & \textbf{17.28} & \textbf{13.62} & \textbf{14.23} & \textbf{11.16} & \textbf{15.05} & \textbf{11.82} & \textbf{14.77} & \textbf{11.48} & \textbf{14.13} & \textbf{11.02} & \textbf{10.58} & \textbf{7.78} & \textbf{16.01} & \textbf{12.53} \\ \Xhline{2\arrayrulewidth}

\multirow{10}{*}{\begin{tabular}[c]{@{}c@{}}Amazon\\ Elec.\end{tabular}} & SSRM & 7.33 & 5.13 & 6.82 & 4.86 & 7.82 & 5.62 & 7.13 & 5.07 & 7.12 & 5.19 & 6.84 & 5.08 & 7.39 & 5.39 & 6.29 & 4.65 & 7.08 & 5.12 \\  

& GAG & 9.47 & 7.37 & 8.64 & 6.56 & 8.85 & 6.62 & 8.07 & 6.05 & 7.89 & 6.03 & 7.79 & 5.98 & 8.16 & 6.26 & 7.41 & 5.47 & 8.28 & 6.29 \\ 

& GIUA-GNN & 11.63 & 9.74 & 10.57 & 8.73 & 10.30 & 8.23 & 9.22 & 7.37 & 8.87 & 7.10 & 8.95 & 7.31 & 9.40 & 7.69 & 8.10 & 6.42 & 9.63 & 7.84 \\ 

& $\text{MMLP}_{\text{RH}}$ & 10.08 & 8.07 & 9.14 & 7.19 & 9.39 & 7.14 & 8.34 & 6.43 & 7.77 & 6.17 & 7.60 & 6.12 & 8.14 & 6.60 & 6.88 & 5.50 & 8.42 & 6.65 \\ 

& $\text{LGMRec}_{\text{RH}}$ & 9.65 & 7.99 & 8.84 & 6.97 & 8.24 & 6.41 & 7.83 & 5.64 & 6.99 & 5.26 & 7.03 & 5.76 & 7.59 & 5.78 & 6.21 & 5.08 & 7.80 & 6.11 \\ 

& $\text{IISAN}_{\text{RH}}$ & 16.55 & 13.26 & \textbf{15.50} & \textbf{12.08} & 14.88 & 11.35 & 13.59 & 10.35 & 12.39 & 9.35 & 12.41 & 9.32 & 12.99 & 9.46 & 10.95 & 7.95 & 13.66 & 10.39 \\ 

& $\text{MMLP}_{\text{KD}}$ & 10.07 & 8.07 & 9.15 & 7.19 & 9.31 & 7.07 & 8.17 & 6.28 & 7.67 & 6.12 & 7.36 & 5.73 & 8.09 & 6.48 & 6.78 & 5.45 & 8.33 & 6.55 \\ 

& $\text{LGMRec}_{\text{KD}}$ & 9.63 & 7.97 &  8.77 & 6.97 & 8.26 & 6.42 & 7.67 & 5.48 & 6.26 & 4.87 & 6.98 & 5.56 & 7.40 & 5.91 & 6.17 & 5.06 & 7.64 & 6.03 \\  

& $\text{IISAN}_{\text{KD}}$ & 15.27 & 12.18 & 14.14 & 11.07 & 13.38 & 10.28 & 12.27 &  9.33 & 11.11 & 8.50 & 10.91 & 8.36 & 11.28 & 8.58 & 9.18 & 6.77 & 12.19 & 9.29 \\ 

& XSMoE & \textbf{17.07} & \textbf{13.39} & 15.49 & 11.84 & \textbf{15.04} & \textbf{11.39} & \textbf{13.74} & \textbf{10.40} & \textbf{12.67} & \textbf{9.54} & \textbf{12.60} & \textbf{9.41} & \textbf{13.11} & \textbf{9.71} & \textbf{11.32} & \textbf{8.19} & \textbf{13.88} & \textbf{10.48} \\ \Xhline{2\arrayrulewidth}

\multirow{10}{*}{H\&M} & SSRM & 1.02 & 0.50 & 1.62 & 0.83 & 2.14 & 1.14 & 2.51 & 1.26 & 2.54 & 1.37 & 2.06 & 1.10 & 2.60 & 1.42 & 2.29 & 1.29 & 2.10 & 1.11 \\  

& GAG & 1.74 & 0.94 & 2.69 & 1.52 & 3.47 & 2.01 & 4.06 & 2.33 & 3.96 & 2.31 & 3.67 & 2.20 & 3.97 & 2.36 & 3.12 & 1.85 & 3.33 & 1.95 \\ 

& GIUA-GNN & 4.35 & 2.81 & 5.61 & 3.72 & 5.94 & 3.77 & 7.02 & 4.56 & 6.39 & 4.15 & 6.03 & 3.91 & 6.11 & 4.00 & 5.30 & 3.51 & 5.85 & 3.80 \\ 
 
& $\text{MMMLP}_{\text{RH}}$ & 10.31 & 8.18 & 13.36 & 10.75 & 12.23 & 10.07 & 13.85 & 11.13 & 13.72 & 11.01 & 12.99 & 10.41 & 12.37 & 9.77 & 10.14 & 8.06 & 12.37 & 9.92\\

& $\text{LGMRec}_{\text{RH}}$ & 9.73 & 7.66 & 12.71 & 9.94 & 10.47 & 8.30 & 12.56 & 9.73 & 12.28 & 9.60 & 11.74 & 9.16 & 11.93 & 9.41 & 8.96 & 6.96 & 11.30 & 8.84 \\ 

& $\text{IISAN}_{\text{RH}}$ & 26.22 & 22.09 & 27.80 & 23.97 & 25.50 & 20.27 & 24.71 & 19.18 & 23.06 & 19.01 & 22.59 & 17.61 & 24.09 & 19.04 & 25.57 & 20.49 & 24.94 & 20.21 \\

& $\text{MMMLP}_{\text{KD}}$ & 10.30 & 8.15 & 13.32 & 10.73 & 12.19 & 10.03 & 13.85 & 11.06 & 13.59 & 10.9 & 13.03 & 10.42 & 12.18 & 9.53 & 9.97 & 8.00 & 12.30 & 9.86 \\ 

& $\text{LGMRec}_{\text{KD}}$ & 9.71 & 7.66 &  12.66 & 9.90 & 10.45 & 8.25 & 12.06 & 9.29 & 12.44 & 9.71 & 11.81 & 9.18 & 11.23 & 8.62 & 9.14 & 7.15 & 11.19 & 8.72 \\ 

& $\text{IISAN}_{\text{KD}}$ & 25.29 & 20.33 & 27.64 & 22.82 & 24.29 & 19.40 &23.09 & 18.08 & 22.79 & 18.16 & 21.66 & 16.93 & 23.22 & 18.28 & 24.67 & 19.73 & 24.08 & 19.22 \\

& XSMoE & \textbf{26.77} & \textbf{22.39} & \textbf{28.98} & \textbf{24.17} & \textbf{25.76} & \textbf{20.47} & \textbf{24.86} & \textbf{19.22} & \textbf{24.06} & \textbf{19.13} & \textbf{22.88} & \textbf{17.75} & \textbf{24.43} & \textbf{19.19} & \textbf{25.81} & \textbf{20.74} & \textbf{25.44} & \textbf{20.38} \\ \Xhline{2\arrayrulewidth}

\end{tabular}%
}
\label{tab:overall}
\end{table*}

\subsubsection{Evaluation Protocol}
We follow the evaluation protocols as in previous streaming recommendation literature \cite{qiu2020gag, guo2019streaming, yin2023multi}. We begin by sorting the entire dataset chronologically and dividing it into 10 mutually exclusive sequential chunks: \([D_0, D_1, \dots, D_{9}]\). Each chunk has an equal number of records. The model is initially warmed up using data from \(D_0\). Then, for each time window \(s \in [1, 8]\), the model is updated on \(D_s\) and evaluated on the subsequent chunk \(D_{s+1}\). We assess performance using two standard metrics in recommendation research: Hit Rate at 10 (HR@10) and Normalized Discounted Cumulative Gain at 10 (NDCG@10), following previous work \cite{qu2024budgeted, qu2023continuous, zhang2024modeling, zhang2024qagcf, qu2024sparser, zhang2025testtime, interaction2023yuan, zhang2024comprehensive}.

\subsection{Implementation Details}
We now present the implementation details of XSMoE. We use ``bert-base-uncased" as the textual encoder and ``vit-base-patch16-224" as the visual encoder. In each time window, training is monitored using NDCG@10 on the validation set, with early stopping triggered if performance does not improve for 5 consecutive epochs. We adopt the Adam optimizer \cite{kingma2014Adam} with an initial learning rate of 0.001, which decays by a factor of 0.95 after each epoch, down to a minimum of 0.0001. At the beginning of each new time window, the learning rate is reset to 0.001. The batch size is 256. To further minimize computational overhead, for BERT and ViT, we group the first six Transformer layers together and the latter six layers together, resulting in two layers in each side-tuning network. The hidden size of the experts $d'$ is 64. For the item sequence encoder, the number of Transformer blocks is 2. The maximal sequence length is 10. The maximal number of words in texts is 40. Half-precision training is used to the proposed method as well as all baselines.

\subsection{Overall Performance Comparison}
Table \ref{tab:overall} reports the recommendation performance w.r.t. HR@10 and NDCG@10 across different time windows and datasets. Our proposed method, XSMoE, achieves the best performance on all three datasets in nearly all time windows. We run XSMoE and IISAN, which consistently achieves the second best performance, 5 times and perform student t-test on the average HR@10 and NDCG@10 to further verify the efficacy of XSMoE. With all the $p$-values being less than 0.01, we can confidently conclude that the advantageous performance of XSMoE is statistically significant and did not occur by chance, thus answering RQ1.

We observe that ID-based method GIUA-GNN outperforms MMMLP and LGMRec on the two Amazon datasets, which shows that multimodal recommenders that simply extract multimodal features without fine-tuning the multimodal encoders do not necessarily achieve better performance than traditional ID-based models, consistent with the findings in \cite{ni2023content}. It is also evident that IISAN and XSMoE outperform MMMLP and LGMRec by a large margin on all the three datasets. 

By comparing the performance of the three static multimodal recommenders paired with reservoir sampling (i.e., $\text{MMMLP}_{\text{RH}}$, $\text{LGMRec}_{\text{RH}}$, $\text{IISAN}_{\text{RH}}$) and KD loss (i.e., $\text{MMMLP}_{\text{KD}}$, $\text{LGMRec}_{\text{KD}}$, $\text{IISAN}_{\text{KD}}$), we can see reservoir sampling achieves better results than KD loss. This suggests that while KD loss attempts to preserve past knowledge by constraining parameter updates, it is not necessarily useful to the current task and may limit the model's ability to adapt to new data in SRSs, consistent with the conclusion in \cite{zhang2023incremental}. 

A general decline in performance can be seen over time, particularly on the two Amazon datasets. This degradation is primarily due to the evaluation setting: when testing on \( D_s \), the model ranks all items seen from \( [D_0, D_1, \ldots, D_s] \). In realistic e-commerce scenarios, however, many older items become obsolete and are no longer available. For example, a product listed in 1996 is unlikely to remain purchasable in 2025, even if a user would still have liked it if it remained on the market. Our evaluation protocol does not consider item obsolescence and includes all historical items during ranking. As the Amazon datasets span a long time period (1996–2018), the accumulation of obsolete items contributes to a more noticeable decline in performance over time compared to the H\&M dataset that only includes data from 2022.

\subsection{Ablation Studies and Sensitivity Analysis} \label{sec:ablation}

\begin{table*}[t] 
\caption{Recommendation performance w.r.t. HR@10 and NDCG@10 in percentage (\%) of different variations of XSMoE on the Amazon Home, Amazon Electronics (abbreviated as Amazon Elec.), and H\&M datasets. $[D_2, \dots, D_9]$ denote the test data chunks. For XSMoE, $\tau$ is set to 0.15, 0, and 0 on the Amazon Home, Amazon Electronics, and H\&M datasets, respectively. The best results are highlighted.}
\label{tab:ablation}
\resizebox{\textwidth}{!}{%
\begin{tabular}{cccccccccccccccccccc}
\Xhline{2\arrayrulewidth}

\multirow{2}{*}{Datasets} & \multirow{2}{*}{Variants} & \multicolumn{2}{c}{$D_2$} & \multicolumn{2}{c}{$D_3$} & \multicolumn{2}{c}{$D_4$} & \multicolumn{2}{c}{$D_5$} & \multicolumn{2}{c}{$D_6$} & \multicolumn{2}{c}{$D_7$} & \multicolumn{2}{c}{$D_8$} & \multicolumn{2}{c}{$D_9$} & \multicolumn{2}{c}{AVG} \\ \cline{3-20}

& & HR & NDCG & HR & NDCG & HR & NDCG & HR & NDCG & HR & NDCG & HR & NDCG & HR & NDCG & HR & NDCG & HR & NDCG \\ \cline{1-20}

\multirow{5}{*}{\begin{tabular}[c]{@{}c@{}}Amazon\\ Home\end{tabular}} & Visual & 21.08 & 17.41 & 17.91 & 14.70 & 15.74 & 12.91 & 12.78 & 10.46 & 13.54 & 11.10 & 13.32 & 10.85 & 12.52 & 10.24 & 9.00& 7.09 & 15.13 & 12.42 \\ 

& Textual & 21.26 & 17.21 & 17.92 & 14.51 & 15.69 & 12.77 & 12.72 & 10.26 & 13.37 & 10.81 & 13.35 & 10.64 & 12.59 & 10.09 & 9.15 & 7.07 & 15.26 & 12.31 \\ 

& Static & 22.00 & 17.45 & 18.95 & 15.00 & 16.99 & 13.53 & 14.02 & 11.00 & 14.65 & 11.59 & 14.41 & 11.32 & 13.68 & 10.72 & 10.37 & 7.72 & 15.63 & 12.29 \\ 

& NoFT & 21.03 & 16.92 & 18.23 & 14.59 & 16.12 & 13.02 & 13.29 & 10.67 & 13.78 & 11.14 & 13.78 & 11.06 & 12.96 & 10.39 & 9.55 & 7.38 & 14.84 & 11.90 \\ 

& XSMoE & \textbf{22.67} & \textbf{18.02} & \textbf{19.43} & \textbf{15.37} & \textbf{17.28} & \textbf{13.62} & \textbf{14.23} & \textbf{11.16} & \textbf{15.05} & \textbf{11.82} & \textbf{14.77} & \textbf{11.48} & \textbf{14.13} & \textbf{11.02} & \textbf{10.58} & \textbf{7.78} & \textbf{16.01} & \textbf{12.53} \\ \Xhline{2\arrayrulewidth}

\multirow{5}{*}{\begin{tabular}[c]{@{}c@{}}Amazon\\ 
Elec.\end{tabular}} & Visual & 15.52 & 12.67 & 14.13 & 11.48 & 13.12 & 10.39 & 12.12 & 9.58 & 11.06 & 8.69 & 10.76 & 8.48 & 11.13 & 8.81 &  8.93 & 6.89 & 12.43 & 10.00 \\ 

& Textual & 15.26 & 12.49 & 14.26 & 11.23 & 12.90 & 10.18 & 12.02 & 9.47 & 10.74 & 8.36 & 10.80 & 8.52 & 11.23 & 8.74 & 9.31 & 7.07 & 12.46 & 9.91 \\ 

& Static & 16.57 & 13.06 & 15.29 & 11.75 & 14.84 & 11.22 & 13.54 &  10.20 & 12.45 & 9.31 & 12.41 & 9.30 & 13.00 & \textbf{9.71} & 10.97 & 7.97 & 13.63 & 10.32 \\ 

& NoFT & 16.09 & 12.89 & 14.67 & 11.59 & 14.06 & 10.79 & 12.84 & 9.91 & 11.72 & 9.03 & 11.36 & 8.86 & 11.90 & 9.13 & 9.94 & 7.42 & 12.82 & 9.95 \\ 

& XSMoE & \textbf{17.07} & \textbf{13.39} & \textbf{15.49} & \textbf{11.84} & \textbf{15.04} & \textbf{11.39} & \textbf{13.74} & \textbf{10.40} & \textbf{12.67} & \textbf{9.54} & \textbf{12.60} & \textbf{9.41} & \textbf{13.11} & \textbf{9.71} & \textbf{11.32} & \textbf{8.19} & \textbf{13.88} & \textbf{10.48} \\ \Xhline{2\arrayrulewidth}

\multirow{5}{*}{H\&M} & Visual & 22.53 & 20.62 & 24.67 & 22.52 & 21.05 & 18.31 & 20.63 & 17.31 & 20.39 & 17.62 & 18.99 & 16.22 & 20.27 & 17.47 & 21.26 & 18.81 & 21.32 & 18.76 \\ 

& Textual & 25.27 & 17.44 & 26.27 & 17.87 & 22.72 & 14.93 & 21.28 & 13.99 & 21.47 & 14.21 & 20.44 & 13.49 & 22.10 & 14.93 & 24.2 & 16.72 & 23.17 & 15.65 \\ 

& Static & 25.33 & 21.73 & 27.35 & 23.15 & 24.26 & 19.54 & 23.99 & 18.80 & 23.49 & 18.65 & 22.11 & 17.40 & 23.82 & 18.85 & 25.07 & 19.82 & 24.43 & 19.74 \\ 

& NoFT & 24.38 & 21.43 & 26.76 & 23.45 & 23.39 & 19.39 & 22.79 & 18.35 & 22.27 & 18.45 & 21.15 & 17.20 & 22.63 & 18.57 & 24.13 & 20.26 & 23.44 & 19.64 \\ 

& XSMoE & \textbf{26.77} & \textbf{22.39} & \textbf{28.98} & \textbf{24.17} & \textbf{25.76} & \textbf{20.47} & \textbf{24.86} & \textbf{19.22} & \textbf{24.06} & \textbf{19.13} & \textbf{22.88} & \textbf{17.75} & \textbf{24.43} & \textbf{19.19} & \textbf{25.81} & \textbf{20.74} & \textbf{25.44} & \textbf{20.38} \\ \Xhline{2\arrayrulewidth}

\end{tabular}}

\end{table*}

\subsubsection{Expandable Side MoE} 
To preserve previously learned knowledge and mitigate catastrophic forgetting, XSMoE uses the MoE architecture that incrementally allocates new expert networks to learn new data. To assess the effectiveness of this architecture, we disable the MoE architecture and only use one trainable FFN expert per layer inside the side-tuning network throughout the training process. We denote this static variant as ``Static". As shown in Table \ref{tab:ablation}, XSMoE consistently outperforms Static, highlighting the efficacy of the expandable MoE architecture in multimodal SRSs and thus answering RQ3.

\begin{table}[t] 
\caption{Comparison of efficiency across total model parameters (MP), trainable parameters (TrP), GPU memory usage (Memory), total training time (TT), and average training time per epoch (ET) on the Amazon Home, Amazon Electronics (abbreviated as Amazon Elec.), and H\&M datasets. MP, TrP, and Memory are measured in MB, while TT and ET are measured in seconds.}
\resizebox{0.8\columnwidth}{!}{ 
\label{tab:efficiency}
\begin{tabular}{ccccccc}
\Xhline{2\arrayrulewidth}
 Datasets & Variants & MP & TrP & Memory & TT & ET \\ \cline{1-7} 
 \multirow{7}{*}{\begin{tabular}[c]{@{}c@{}}Amazon\\ Home\end{tabular}} & NoFT & 2.27 &  2.27 & 600.96 & 5694 & 24.33 \\ 
 & $\tau=0$ & 15.88 & 3.88 & 732.78 & 3514 & 24.07 \\  
 & $\tau=0.05$ & 10.21 & 3.84 & 620.01 & 3425 & 23.95 \\  
 & $\tau=0.10$ & 8.32 & 3.82 & 616.20 & 3246 & 23.69 \\ 
 & $\tau=0.15$ & 7.95 & 3.82 & 615.44 & 3292 & 24.03 \\ 
 & $\tau=0.20$ & 6.43 & 3.81 & 612.39 & 3557 & 23.87 \\ 
 & $\tau=0.25$ & 6.06 & 3.81 & 611.63 & 3820 & 24.64 \\  
 \Xhline{2\arrayrulewidth}

\multirow{7}{*}{\begin{tabular}[c]{@{}c@{}}Amazon\\ Elec.\end{tabular}} 
 & NoFT & 2.27 & 2.27 & 577.99 & 6472 & 24.80 \\  
 & $\tau=0$ & 15.88 & 3.88 & 732.78 & 3782 & 25.38  \\ 
 & $\tau=0.05$ & 10.21 & 3.84 & 596.23 & 3741 & 24.94 \\ 
 & $\tau=0.10$ & 9.46 & 3.83 & 594.70 & 3888 & 24.92 \\ 
 & $\tau=0.15$ & 8.32 & 3.82 & 592.42 & 4005 & 24.72 \\ 
 & $\tau=0.20$ & 6.81 & 3.81 & 589.37 & 4216 & 25.09 \\ 
 & $\tau=0.25$ & 6.81 & 3.81 & 589.37 & 4139 & 25.09 \\ 
\Xhline{2\arrayrulewidth}

\multirow{8}{*}{H\&M}  
 & NoFT & 2.27 &  2.27 & 274.97 & 4477 & 20.92  \\ 
 & $\tau=0$ & 15.88 & 3.88 & 732.78 & 2846 & 20.93 \\ 
 & $\tau=0.05$ & 10.59 & 3.84 & 533.07 & 2778 & 20.58 \\ 
 & $\tau=0.10$ & 8.70 & 3.83 & 453.15 & 2765 & 20.63 \\ 
 & $\tau=0.15$ & 7.57 & 3.82 & 423.32 & 2924 & 21.50 \\ 
 & $\tau=0.20$ & 6.43 & 3.81 & 393.94 & 2972 & 20.93 \\ 
 & $\tau=0.25$ & 6.06 & 3.81 & 384.88 & 2749 & 20.36 \\ \Xhline{2\arrayrulewidth}
\end{tabular}%
}
\end{table}

\subsubsection{Expert Pruning} 
As discussed in Section~\ref{sec:pruning}, to prevent unbounded model growth and complexity, we measure the utilization of each expert network at the end of every training epoch using a norm-based heuristic. Experts whose utilization falls below a predefined threshold $\tau$ are pruned at the end of each time window. We evaluate pruning sensitivity by searching over $\tau \in \{0, 0.05, 0.10, 0.15, 0.20, 0.25\}$. Setting $\tau=0$ effectively disables pruning, while a larger $\tau$ result in more aggressive pruning.

\begin{figure}[h]
    \begin{subfigure}{\columnwidth}
        \caption*{Amazon Home}
        \includegraphics[width=0.8\columnwidth]{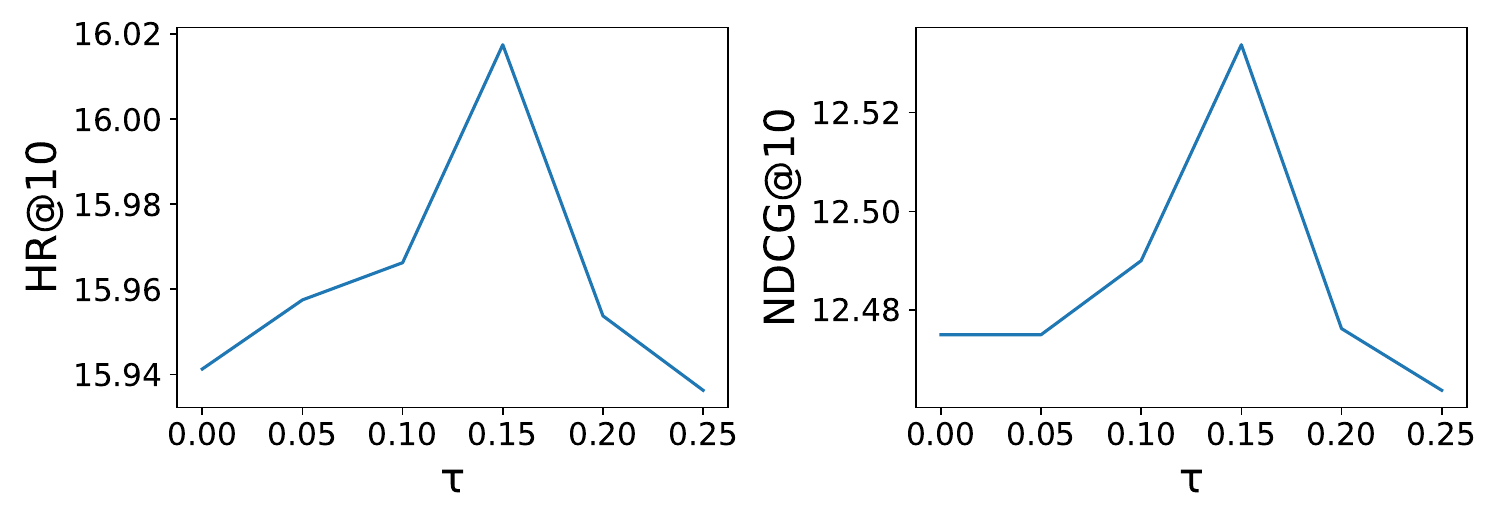}
    \end{subfigure}
    \vspace{-1em}
    \begin{subfigure}{\columnwidth}
        \caption*{Amazon Electronics}
        \includegraphics[width=0.8\columnwidth]{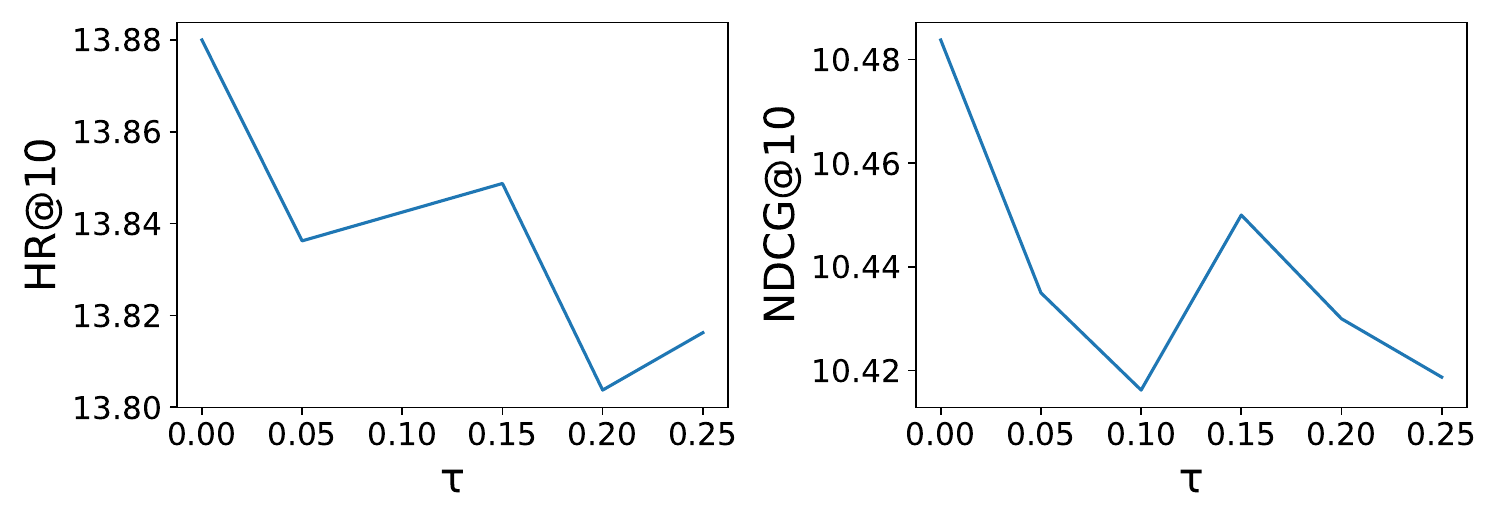}
    \end{subfigure}
    \vspace{-1em}
    \begin{subfigure}{\columnwidth}
        \caption*{H\&M}
        \includegraphics[width=0.8\columnwidth]{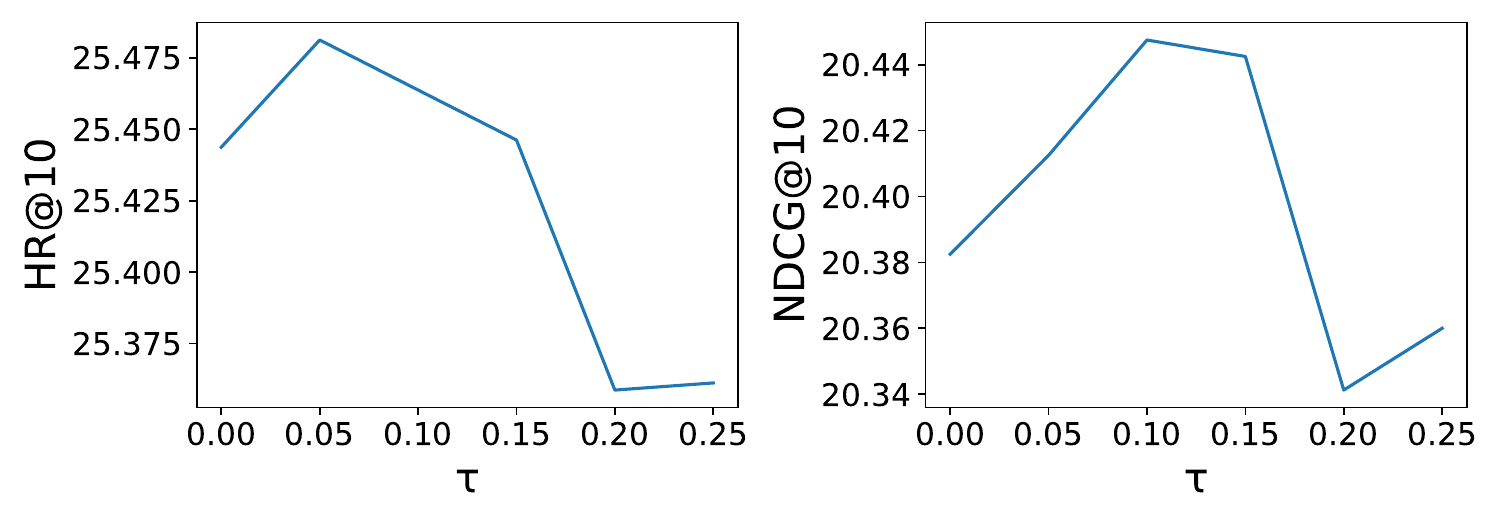}
    \end{subfigure}
    \caption{Sensitivity analysis of the hyperparameter $\tau$ w.r.t. HR@10 and NDCG@10 on the Amazon Home, Amazon Electronics, and H\&M datasets.}
    \label{fig:tau}
\end{figure}

\textbf{Recommendation performance}. Figure~\ref{fig:tau} presents the recommendation performance across different $\tau$ values in terms of HR@10 and NDCG@10, while Table~\ref{tab:efficiency} reports the corresponding efficiency metrics, including total and trainable parameters, GPU memory usage, total training time, and average per-epoch time. As shown in Figure~\ref{fig:tau}, expert pruning causes only a marginal drop in recommendation performance on the Amazon Electronics dataset while a slight increase is witnessed on the Amazon Home and H\&M datasets with $\tau \in [0.05, 0.15]$.

\textbf{Efficiency}. As shown in Table~\ref{tab:efficiency}, increasing the pruning threshold $\tau$ initially results in a sharp reduction in total parameters and GPU memory usage. However, this downward trend plateaus once $\tau > 0.10$. While per-epoch training time remains nearly constant across different $\tau$ values, overall wall-clock time is minimized for moderate pruning thresholds ($\tau \in [0.05, 0.10]$), as the smaller models converge more quickly. When $\tau$ exceeds 0.10, total training time increases again, suggesting that excessive pruning diminishes model expressiveness and slows convergence. In summary, a moderate pruning threshold offers significant gains in model parameter efficiency, memory efficiency, and overall training speed without sacrificing recommendation performance, hence answering RQ4.

\subsubsection{Finetuning}
To evaluate recommendation performance without any fine-tuning, we remove the entire side-tuning network and use the pretrained backbone encoders directly to extract multimodal features without fine-tuning. We refer to this variant as ``NoFT". Its recommendation performance and efficiency are summarized in Tables~\ref{tab:ablation} and~\ref{tab:efficiency}, respectively. As shown in Table~\ref{tab:ablation}, XSMoE consistently outperforms NoFT, demonstrating the importance of fine-tuning for high-quality recommendations and addressing RQ2. Despite having more trainable parameters, XSMoE with various $\tau$ values achieves comparable per-epoch training times to NoFT (Table~\ref{tab:efficiency}). More importantly, NoFT requires substantially more epochs to converge, indicating that lower parameter count does not necessarily lead to faster training. Instead, XSMoE’s adaptive side-tuning improves model expressiveness, enabling faster training.

\subsubsection{Importance of Each Modality}
XSMoE leverages both image and text modalities for item representation. To assess the contribution of each modality to recommendation performance and answer RQ5, we conduct an ablation study by removing one modality at a time and evaluating the model using HR@10 and NDCG@10. We denote the two XSMoE variations that only utilize the visual and textual modality as ``Visual'' and ``Textual'', respectively. Their recommendation performance w.r.t. HR@10 and NDCG@10 is presented in Table \ref{tab:ablation}. As the results show, the recommendation performance decreases significantly after the removal of either modality. For the two Amazon datasets, model with only textual and only visual modality achieve similar performance. For the H\&M dataset, however, the model with only textual modality witnesses better HR@10 compared to the visual-only model while experiencing a sharp drop in NDCG@10.

\section{Conclusion and Future Research}
In this work, we introduced Expandable Side Mixture-of-Experts (XSMoE), a novel framework for multimodal streaming recommendation that addresses the twin challenges of efficient adaptation and continual knowledge retention. By leveraging side-tuning for memory-efficient training, dynamically expanding a Mixture-of-Experts network to encode new information, and pruning underutilized experts to control model growth, XSMoE achieves both computational efficiency and strong performance. Our design integrates frozen pretrained visual and textual encoders with lightweight, modular experts that adapt over time without overwriting past knowledge. Extensive experiments demonstrate that XSMoE outperforms state-of-the-art streaming recommender models, offering a practical solution for real-time multimodal recommendation in streaming environments. In future work, we aim to explore how user/item ID embeddings can be efficiently integrated in multimodal streaming recommendation.

\section{Acknowledgment}
The Australian Research Council partially supports this work under the streams of Future Fellowship (Grant No. FT210100624),  the Discovery Project (Grant No. DP240101108), and the Linkage Projects (Grant No. LP230200892 and LP240200546).

ChatGPT-4o was used for language polishing purposes during the preparation of this manuscript. 

\clearpage
\bibliographystyle{ACM-Reference-Format}
\bibliography{XSMoE}

\end{document}